\documentclass[aps,prb,twocolumn]{revtex4}   
\usepackage{graphicx}

\usepackage{amsmath}   
\usepackage{subfigure} 
\usepackage{amssymb}    
\usepackage{graphicx}   
\usepackage{array}
\usepackage{epstopdf}

\begin{document}
\title{Hybrid deterministic and stochastic approach for efficient long time scale atomistic simulations}
\author{Pratyush Tiwary}
 \email{pt@caltech.edu}   
 \affiliation{Department of Applied Physics and Materials Science, California Institute of Technology, Pasadena, California 91125, USA}
\author{Axel van de Walle}
\affiliation{School of Engineering, Brown University, Providence, Rhode Island 02912, USA}
\date{\today}

\begin{abstract}
We propose a hybrid deterministic and stochastic approach to achieve extended time scales in atomistic simulations that combines the strengths of Molecular Dynamics (MD) and Monte Carlo (MC) simulations in an easy-to-implement way. The method exploits the rare event nature of the dynamics similar to most current accelerated MD approaches but goes beyond them by providing, without any further computational overhead, (a) rapid thermalization between infrequent events, thereby minimizing spurious correlations and (b) control over accuracy of time scale correction, while still providing similar or higher boosts in computational efficiency. We present two applications of the method: (a) vacancy mediated diffusion in Fe yields correct diffusivities over a wide range of temperatures and (b) source controlled plasticity and deformation behavior in Au nanopillars at realistic strain rates ($10^4$/sec and lower) with excellent agreement with previous theoretical predictions and \textit{in situ} high-resolution transmission electron microscopy (HRTEM) observations. The method gives several orders of magnitude improvements in computational efficiency relative to standard MD and good scalability with size of system.

\end{abstract}
\maketitle

With vast improvements in the quality of available interatomic force-fields and computer power, the classical MD simulation has seen a dramatic increase in its use across a variety of fields over the past few decades \cite{md_dislocation,md_bio,proteinfolding,creep}. One of the features that makes MD so appealing is its ability to actually follow the dynamical evolution of the system, thus giving insight into the microscopic behavior of the material. However, this is where the major limitation of MD comes into light too: most of the interesting dynamics occurs as the system moves from one energy basin to another through infrequent rare events, while the system remains stuck in some energy basin for extended periods of time. This non-ergodicity, coupled with the small time steps (on the order of femtoseconds) needed for total energy staying conserved, severely restricts the timescales accessible in MD simulations and also leads to limited phase space exploration.
\begin{figure}[htp]
\begin{center}
 \includegraphics[width=40mm]{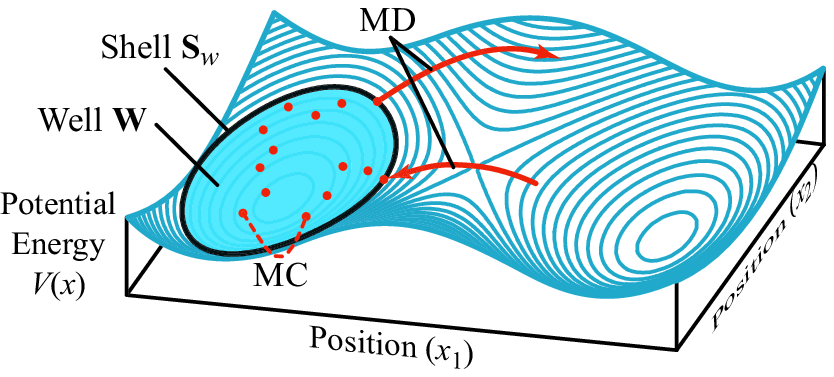}
 \includegraphics[width=40mm]{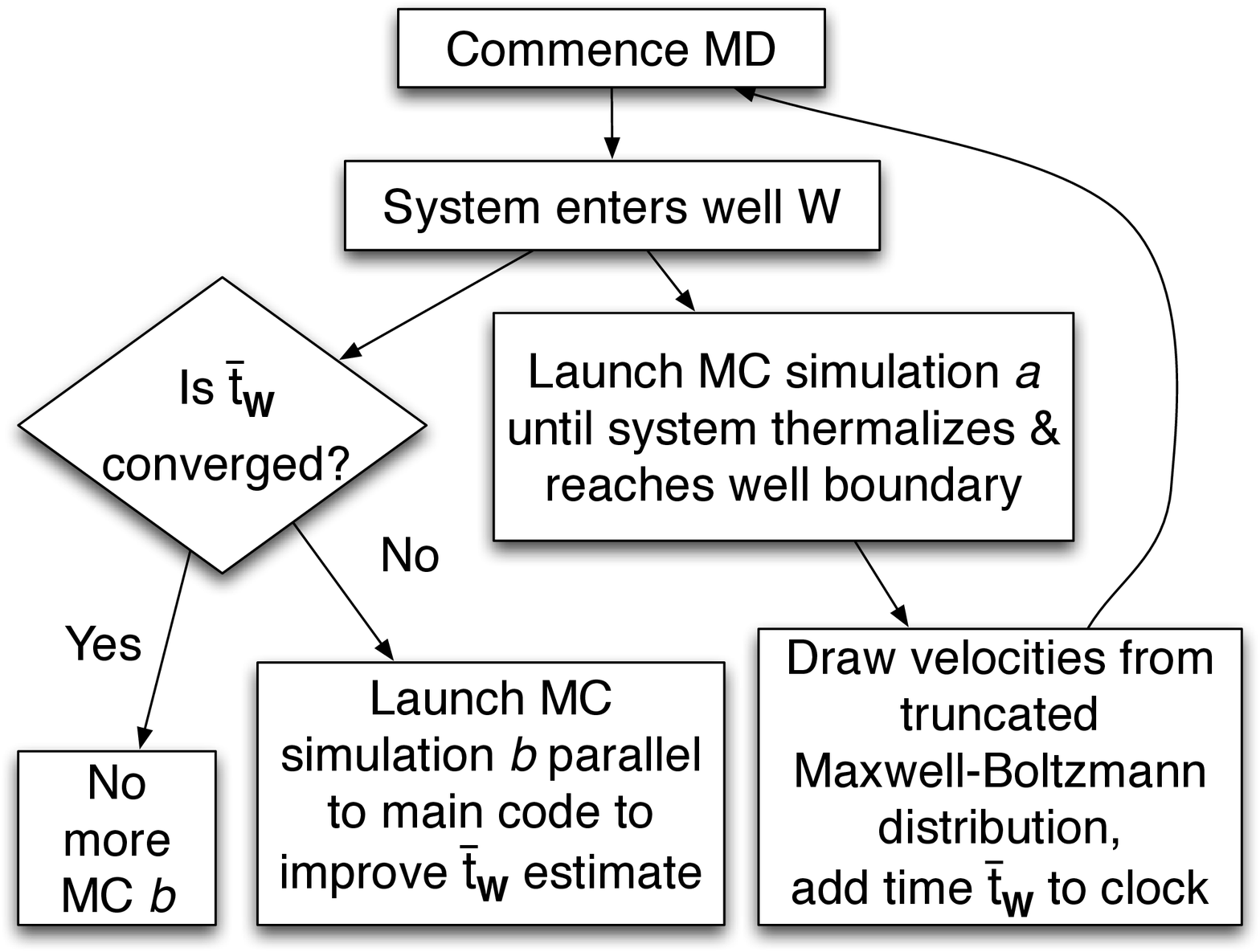}
 \end{center} 
 \caption[1]
{(Color online) Schematic and flowchart of the algorithm. Shell \textbf{S$_w$} of constant potential energy and energy well \textbf{W} as described in text are shown here. See Eq.(\ref{eq:avgt}) for definition of $\overline{t}_\textbf{W}$.}
\label{fig:f1}
\end{figure}
There have been many attempts at addressing this time-scale problem in MD - examples include metadynamics\cite{metadynamics}, the activation relaxation technique\cite{art}, parallel replica dynamics, temperature accelerated dynamics and hyperdynamics\cite{voter_review, voter_prl}. There are several excellent reviews such as Ref. \onlinecite{voter_review} available on the subject. The hyperdynamics method\cite{voter_prl} offers an elegant and practical way to increase the rate of infrequent events. It consists of adding a potential energy bias that makes the potential wells, in which the system normally remains trapped for extended periods, less deep. A time-scale correction is also evaluated in terms of the bias potential. The hyperdynamics method, especially with the advent of a variety of easy to implement biasing forms\cite{hamelberg}, has seen several compelling applications over the past years\cite{hamelberg,fichthorn_prl,juli_dislocation,thinfilm,desorption}. Our approach in this letter builds upon the crucial insights of Voter and co-workers while seeking improvements along two important dimensions. First, it bypasses a fundamental trade-off present in hyperdynamics: a shallower potential well provides faster dynamics but, at the same time, reduces the ability of the modified potential to properly thermalize the system in between the infrequent events, resulting in artificial correlation between these events. Second, our method provides better independent control over the accuracy of time scale correction, while hyperdynamics time scale estimates can remain noisy up to long simulation times, especially for large system sizes (see Ref. \onlinecite{bondboost} for a discussion on this). 
 
Let the state of the system be characterized by position $x$ and velocity $v$, each being a $3N$-dimensional vector for a system of $N$ atoms. When the potential energy $V(x)$ of the system is above a threshold $V_0$, the system evolves via constant-energy (or constant-temperature) MD according to its true Hamiltonian (Fig. \ref{fig:f1}). This high energy region of the phase space contains the interesting but infrequently occuring events. The method is formally correct for any choice of $V_0$; a higher choice of $V_0$ merely limits our ability to monitor the detailed dynamics of some events. When the system's potential energy falls below $V_0$, two MC simulations are initiated (denoted \textit{a} and \textit{b}). Simulation \textit{a} runs MC with a perfectly uniform potential inside the potential well \textbf{W} consisting the points $x$ where the true potential energy $V(x)$ lies below $V_0$ (i.e. all moves are accepted as long as they do not go outside the well). Simulation \textit{a} is run until the system is well thermalized and has lost memory of how it entered the well (this takes a few MC passes, an insignificant amount of wall clock time). MD then resumes with positions drawn from the last MC state that visited the boundary of the potential well.
The vector $v$ of the velocities of all atoms for restarting MD is drawn from a Maxwell-Boltzmann distribution corresponding to the temperature $T$ of interest, conditional on $v \cdot \nabla V(x)>0$ (i.e. we only consider velocities in the half-space pointing outwards of the well).
MC simulation \textit{a} is first of the crucial differences between our approach and hyperdynamics: it ensures proper thermalization of the system between rare events even when using a completely flat potential in the well. Even though it is done with a uniform potential, it does not lead to the molecular structure being completely lost since we rule out all moves that lead to energy higher than V$_0$.

In parallel to simulation \textit{a}, another MC simulation \textit{b} is launched to estimate the mean time the system should have spent in the well \textbf{W}. Akin to simulation \textit{a}, \textit{b} also rejects all moves that land outside the well \textbf{W}. The mean time spent in \textbf{W} is given by the reciprocal of the flux exiting\cite{voter_tst} the well \textbf{W}:
\begin{equation}
\label{eq:rate}
t_\textbf{W} = \lim_{w \rightarrow 0} ( \langle  {\overline{v}\over w} \; 1(x \in \textbf{S$_w$}) \rangle )^{-1}
\end{equation}
where the average $\langle\cdots\rangle$ is taken over $x$ drawn from the well \textbf{W} with a probability density proportional to $e^{-V(x)/(k_B T)}$ where $k_B$ is Boltzmann's constant and the following definitions hold:
$1(A)$ equals 1 if the event $A$ is true and 0 otherwise,
\textbf{S$_w$} is a shell of width $w$ at the boundary of the well \textbf{W}, which can be defined in the limit of small $w$ as
\begin{equation}
\label{eq:lid}
\textbf{S$_w$} = \{x: |V(x)-V_0| \leq w|\nabla V(x)|/2 \}
\end{equation}
and $\overline{v}$ denotes the mean projection of a Maxwell-Boltzmann-distributed velocity along the unit vector $u$ parallel to $\nabla V(x)$, conditional on $v \cdot u>0$. The latter is given by
\begin{equation}
\label{eq:vbar}
\overline{v} =\sqrt{ \frac{k_B T}{2 \pi } \sum_{i=1}^N  \frac{|u_i|^2}{m_i} }
\end{equation}
where $m_i$ is the mass of atom $i$ and $|u_i|$ denotes length of the 3 dimensional subvector of $u$ associated with atom $i$. Note that the Eq.(\ref{eq:vbar}) reduces to the familiar expression\cite{voter_tst} $\overline{v} =\sqrt{k_B T/{2 \pi m}}$ when all atoms have the same mass, in which case $\overline{v}$ factors out of the average in (\ref{eq:rate}). Since Eq.(\ref{eq:rate}) involves an average, it can be approximated by MC simulations. 
However, the most straightforward implementation of this approach would be very inefficient because $x$ would rarely visits the boundary \textbf{S$_w$} of the well. The efficiency can be considerably improved by using a biased potential $V^{*}(x)$
which is the same as the real potential in the high energy regions (i.e. regions outside well \textbf{W} with $V(x) \geq V_0$), but lifted up in the deep energy basins. With this Eq. (\ref{eq:rate}) becomes
\begin{equation}
\label{eq:t_bias}
t_\textbf{W} = \lim_{w \rightarrow 0} \frac{  \langle  e^{-{{\beta}}(V(x)-V^{*}(x))} \rangle^* }
{\langle {\overline{v}\over w} {e^{-{{\beta}}(V(x)-V^{*}(x))} }1(x \in \textbf{S$_w$}) \rangle^* }
\end{equation}
where the averages $\langle\cdots\rangle^*$ are taken over $x$ drawn from the well \textbf{W} with a probability density proportional to $e^{-V^*(x)/(k_B T)}$ and $\beta={1/k_{B}T}$.
MC simulation \textit{b} is the second main difference with hyperdynamics: it provides separate control over the accuracy of the speed up factor since the the length of the MC simulation \textit{b} can be adjusted independently of the length of the whole simulation.

The form of biasing we use is a well established and easy to implement biasing potential used in several implementations of Voter's hyperdynamics method, proposed by Hamelberg \textit{et al.}\cite{hamelberg}:
\begin{eqnarray}
\label{eq:V_bias}
	V^{*}(x)  &=& V(x) 
	+  \left\{ 
\begin{array}{l l}
0 & \quad {V(x) \geq V_{0} } \\
{(V_0 - V(x))^{2}\over{\alpha+V_0 - V(x)}} & \quad {V(x) < V_0}\\ \end{array} \right.
  \end{eqnarray}

The times $t_\textbf{W}$ obtained via MC simulations \textit{b} can be directly added to the physical time spent doing MD simulations to yield the overall physical time of the simulation. However, refinements of the method can yield further improvement in efficiency. Instead of computing $t_\textbf{W}$ for each well \textbf{W}, one may keep a running average 
\begin{equation}
\label{eq:avgt}
\overline{t}_\textbf{W}=\frac{1}{n_\text{b}} \sum_W t_\textbf{W}
\end{equation}
of the time spend in the $n_b$ wells sampled via MC simulation \textit{b} ($n_a$, the number of wells actually visited, may well far exceed $n_b$). Once this average is converged, there is no need to initiate MC simulation \textit{b} anymore. The overall time spent in all the wells will simply be $\overline{t}_\textbf{W}*n_a/n_b$.
Note that there is no need to keep separate averages for different types of wells, which would have been difficult to implement. Although MC simulations \textit{a} still need to be performed for all wells, these converge much more rapidly.
Other efficiency improvements can be obtained by not performing fully converged MC simulations \textit{b} and exploiting the fact that errors will average out over wells in Eq.(\ref{eq:avgt}). Note that this scheme must be used while ensuring that the biasing potential is sufficiently strong so that most of the random errors in Eq.(\ref{eq:t_bias}) are concentrated in the numerator, to avoid a systematic bias due to nonlinearity of the ratio. We would like to point out that only the parameter $w$ is additional to those in any typical hypderdynamics scheme (Hamelberg \textit{et al.}\cite{hamelberg}'s in this case), the choice of which does not effect the result since we extrapolate t$_\textbf{W}$ to the limit of small $w$\cite{voter_tst}. 
\begin{figure}[htp]
\begin{center}
 \includegraphics[width=80mm]{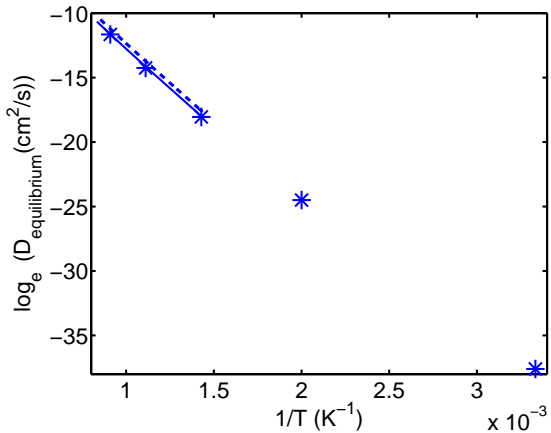}
 \end{center} 
 \caption[2]
{(Color online) Log$_{e}$(diffusivity) versus inverse temperature. (a) Straight lines denote Mendelev \textit{et al.}'s\cite{mendelev} MD calculations. These are valid only until 700 K. (b) Asterisks denote diffusivity measurements per our approach. (c) The dashed line shows experimental measurements \cite{iron_expt} valid between 1000 and 1200 K.}
\label{fig:diffusivity}
\end{figure}
Our approach compares favorably with hyperdynamics\cite{voter_prl} where one does not have control over the accuracy of the accelerated time (hyperdynamics relies on this error cancelling out over time but does not provide an estimate of how much it is\cite{voter_prl,bondboost}), and one is obliged to keep performing dynamics with the biased potential at all stages of the calculation. Thus our method offers boosts as high as those that one could get from setting $\alpha = 0$ in Eq.(\ref{eq:V_bias}) (akin to the flooding scheme\cite{flooding1,flooding2}), but still avoiding the slow convergence in time and problems with discontinuous forces that one encounters in implementing flooding based hyperdynamics. In addition we avoid errors from sampling the system in the state when it is not thermalized between two rare events - once MD is relaunched in our scheme, the system is well thermalized by virtue of simulation \textit{a}. 
To minimize the wall-clock time needed for calculation of time in Eq.(\ref{eq:t_bias}) via simulation \textit{b}, we use an optimal extent of biasing as suggested in Ref. \onlinecite{hamelberg}. This involves setting $\alpha\simeq V_0-V_{min}$ which allows the biased potential to capture the shape of the potential wells\cite{hamelberg}. $\alpha$ smaller than this would improve sampling of the numerator in Eq.(\ref{eq:t_bias}) but detoriate that of the denominator. \begin{figure}[htp]
\begin{center}
 \includegraphics[width=40mm]{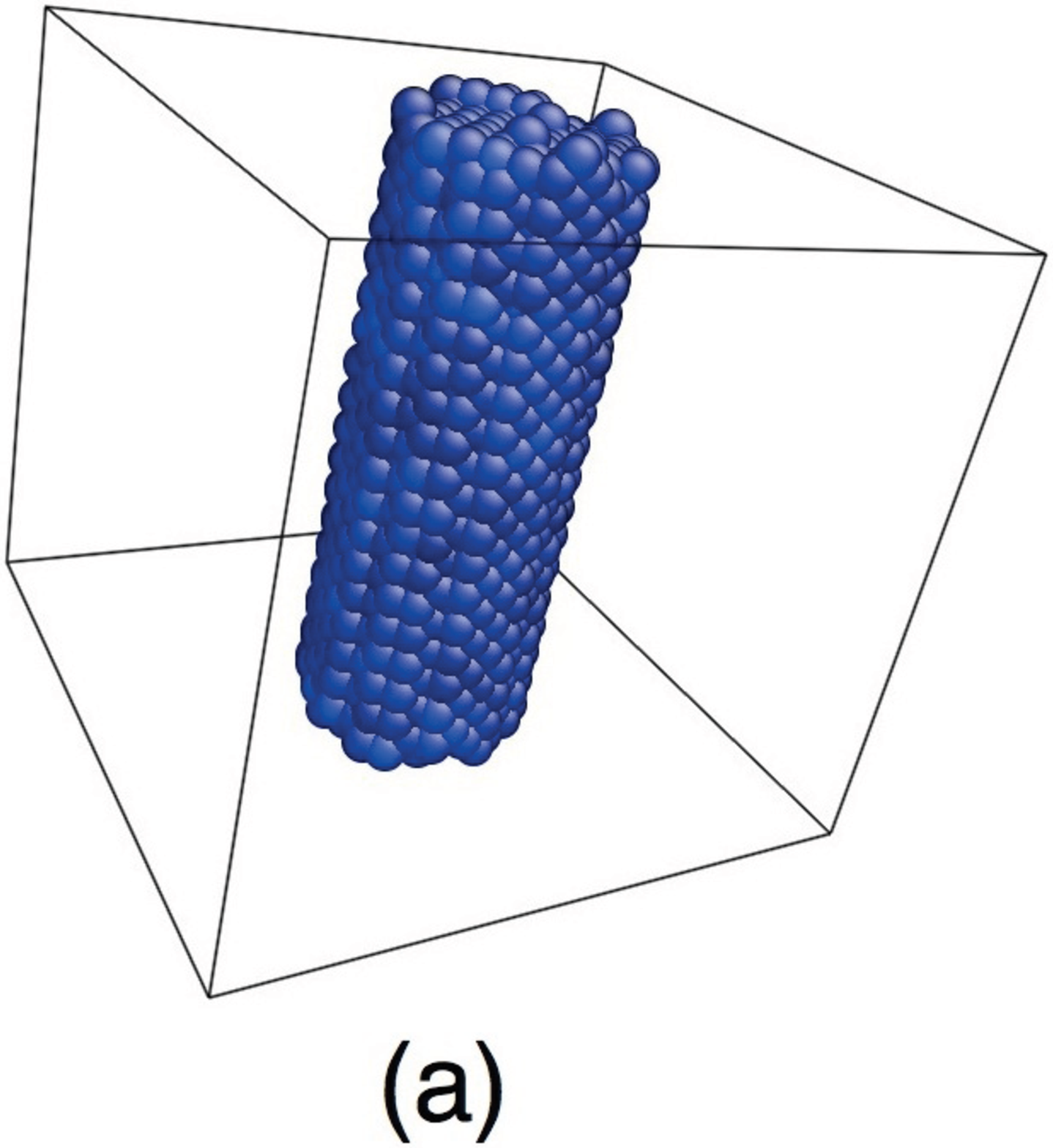}
 \includegraphics[width=40mm]{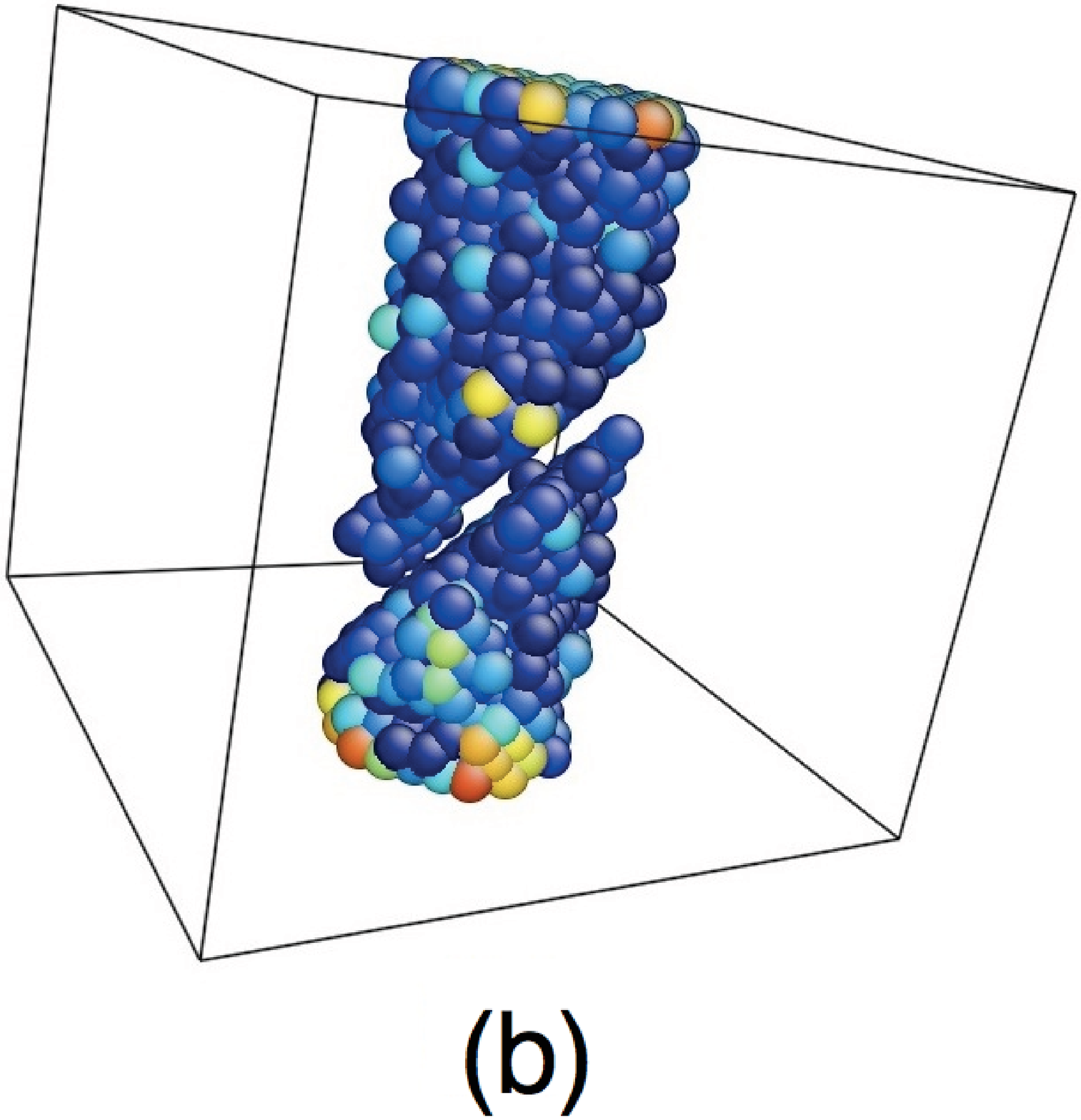} 
 \end{center} 
 \caption[3]
{(Color online) Simulation cell for stress-strain calculations. (a) Prior to application of any strain, (b) after yielding (strain = 12\%) with strain rate = 5x10$^4$/sec. In (b) the leading partial has nucleated on \{111\} slip plane leaving behind the 2-layer thick HCP region denoting an intrinsic stacking fault. Failure is thus through slip and not twinning, in agreement with Ref. \onlinecite{natcomm}. Atoms are identified as per bond order parameter Q$_6$ \cite{steinhardt,atomeye}. Perfect HCP atoms have been removed for clarity.}
\label{fig:cylinder}
\end{figure}
We picked two problems to demonstrate that our method yields correct dynamics: (a) vacancy  diffusion in BCC Fe at room temperature, and (b) deformation behavior in Au nanopillars at realistic strain rates. 

Lattice diffusion at low temperatures is beyond the time scales one can access in current MD simulations, with most investigations\cite{mendelev} only beyond 700 K. The system we consider is 249 Fe atoms (5x5x5 BCC supercell with 1 vacancy) interacting through the Embedded Atom Method (EAM) potential \cite{mendelev}. For the MD part here and in deformation behavior problem, we performed NVT simulations using time step of 2x10$^{-15}$ sec and a Langevin thermostat with coupling constant 1x10$^{-11}$ sec$^{-1}$. The biasing parameter $\alpha$ was 50 eV. The $V_0$ values we used at 500 and 300 K were -975.5 eV and -984 eV resp. (4 and 2.5 eV more than the mean energy at 500 and 300K resp.). We took the equilbrium concentration of defects\cite{mendelev} to convert our effective diffusivity into equilibrium diffusivity. In Fig. \ref{fig:diffusivity} we plot the equilibrium diffusivity as obtained from (a) MD simulations\cite{mendelev},(b) our proposed approach, and (c) experimental measurements\cite{iron_expt} that include contributions from interstitial migrations also and hence are only slightly higher than both ours and MD values. We obtain around 5 orders of magnitude boost, with similar speed up factors for system sizes up to 30000 atoms.

For our second problem (see Fig. \ref{fig:cylinder}), we looked at deformation behavior of Au nanopillars. With advent of excellent \textit{in situ} TEM and HRTEM tools, there are many elegant experiments on sub-10-nm sized crystals\cite{natcomm,bcc,andrew}. Deformation in such small sizes is controlled by dislocation nucleation, and has been phenomenologically predicted\cite{juli_prl} and experimentally found\cite{natcomm,bcc,andrew} to have small activation volumes and strong strain-rate sensitivity. However there is no direct MD based confirmation of this strong strain-rate sensitivity due to inability of MD to reach strain rates lower than $10^7$/sec. 
\begin{figure}[htp]
\begin{center}
 \includegraphics[width=40mm]{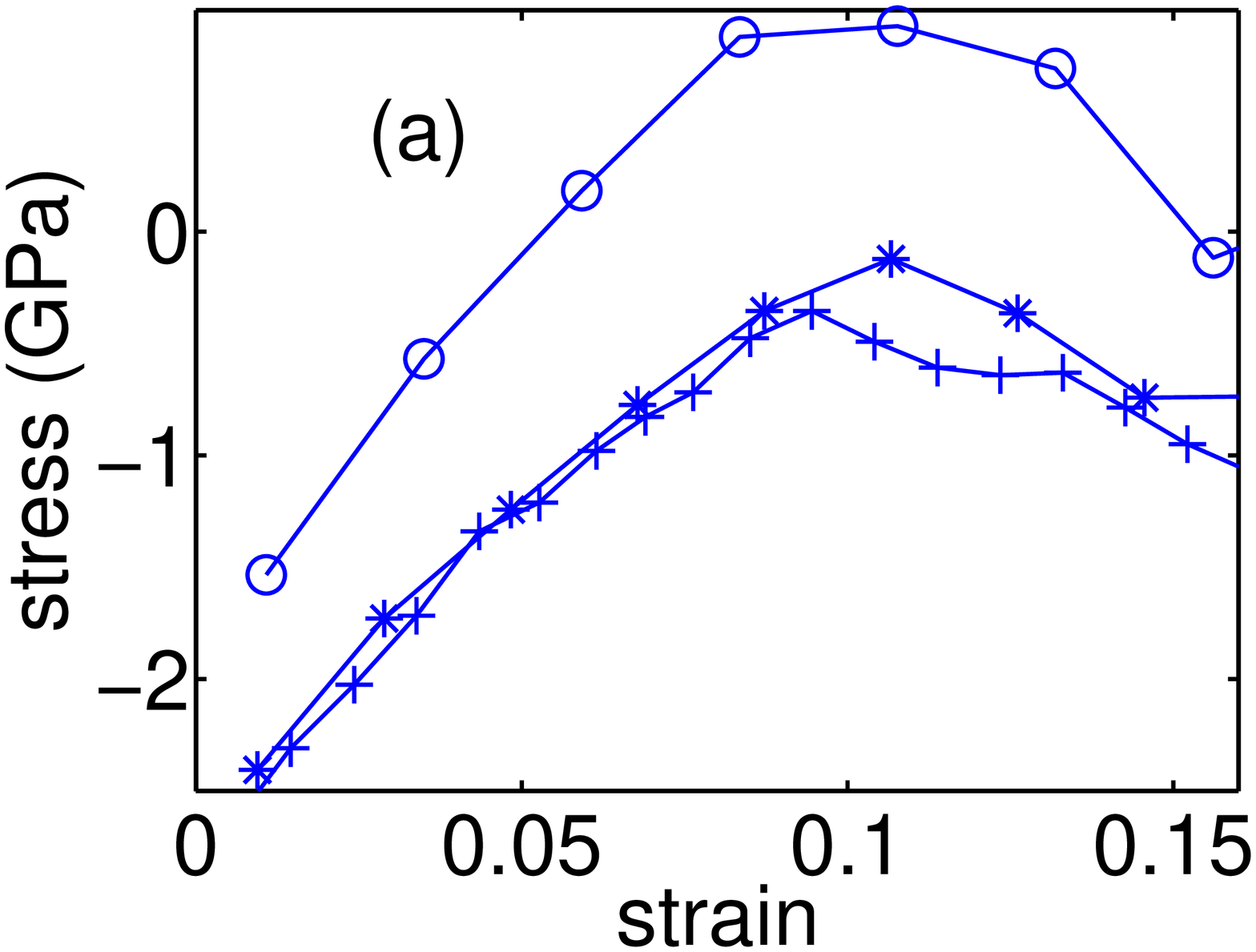}
 \includegraphics[width=40mm]{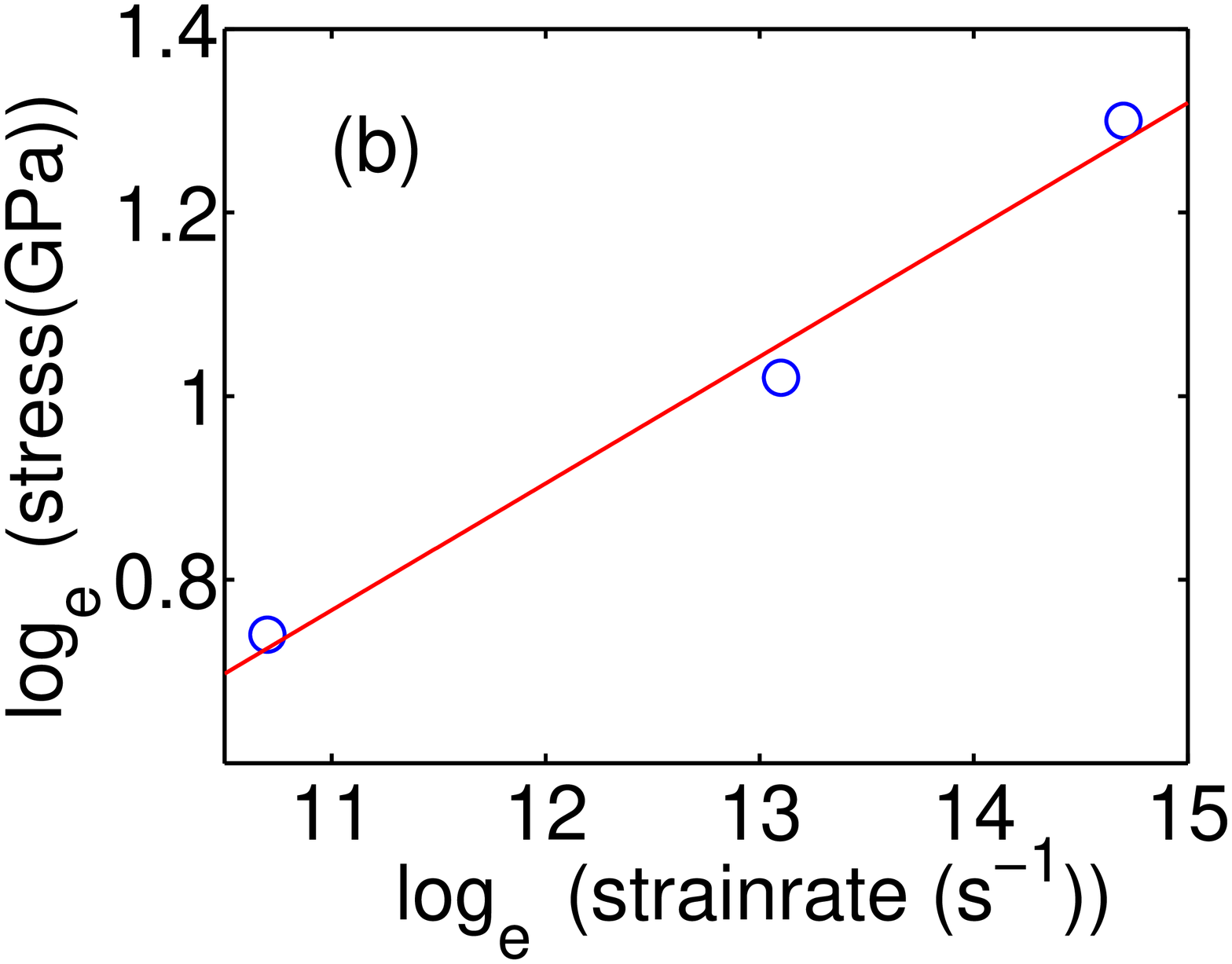}
 \end{center} 
 \caption[4]
{(Color online) (a)Stress-strain plots at 3 different strain rates: 2.5x$10^6$/sec (open circles), 5x$10^5$/sec (asterisks), 5x$10^4$/sec (pluses). The initial stress corresponding to zero-strain is non-zero due to surface effects\cite{srolovitz}.  (b) log$_{e}$(stress) at 11\% strain (relative to surface stress at zero strain) versus log$_{e}$(strain rate).}
\label{fig:stressstrain}
\end{figure} 

Using our method we were able to reach $10^4$/sec strain-rate regime with only around 48 hours of computer time. We could also obtain several correct qualitative and quantitative aspects of the deformation dynamics, without assuming anything about the nature of deformation. The system we consider is 2016 Au atoms (cylinder with height 7.4 nm and diameter 2.5 nm) interacting through EAM potential\cite{grochola}. The biasing parameter $\alpha$ was 1500 eV while the starting $V_0$ value used was -7266 eV(8 eV more than the mean energy at 300K), adjusted every 1000 MD steps to take into account the pressure-volume work contribution from the stress. The cylinder was initially carved out from perfect FCC lattice (Fig. \ref{fig:cylinder}(a)). Periodic boundary conditions were imposed only along the cylinder axis $z$ which is the same as the compression axis $\langle001\rangle$. The cylinder was first equilibrated for 500 ps before beginning the compression carried out by uniformly re-scaling the z-coordinates of all atoms. The atomic virial stress\cite{srolovitz} was used to obtain the Cauchy stress. 4 different strain rates $\dot{\epsilon}$ were considered: 5x$10^6$/sec, 2.5x$10^6$/sec, 5x$10^5$/sec and 5x$10^4$/sec. We present the resulting stress($\sigma$)-strain($\epsilon$) plots in Fig. \ref{fig:stressstrain}(a). Several conclusions can be drawn from Figs. \ref{fig:cylinder} and \ref{fig:stressstrain} that prove our algorithm capable of predicting correct dynamics in complicated systems. The yielding occurs around 10\% strain, and is through slip and not twinning or elastic instabilities: a leading partial nucleates on a \{111\} slip plane at lower stresses than a trailing partial. This can be seen in Fig. \ref{fig:cylinder}(b) where the leading partial nucleated from the surface and left behind a 2-layer thick HCP region which again changes back to FCC after the trailing partial also nucleates at higher stresses and recombines with the leading partial. Fig. \ref{fig:cylinder}(b) is identical to HRTEM images for $\langle001\rangle$ loading of Au nanowires\cite{natcomm}. The strain rate sensitivity $m$ in the relation $\sigma=\sigma_{0}{\dot{\epsilon}}^{m}$ (derived by looking at stress at 11\% strain) is around 0.14$\pm$0.07 (see Fig. \ref{fig:stressstrain}(b), while Ref. \onlinecite{andrew} reports it to be around 0.11 for 75 nm diameter pillars. The activation volume $\Omega$ for the deformation as calculated through\cite{andrew} $\Omega = \sqrt{3}k_B T\partial({lg\dot{\epsilon}})/\partial{\sigma}$is around 1$b^3$ ($b$=burgers vector) in excellent agreement with experiments observations\cite{juli_prl,andrew}. 

To summarize, we have proposed an approach that combines the strengths of MC and MD, thus offering boosts of several orders of magnitudes with good system size scaling. We have applied the method to study lattice diffusion in BCC Fe at low temperatures and deformation of Au nanopillars at low strain rates and found it to work really well in both cases, predicting correct dynamics and exhibiting good scaling with increase in system size from 249 to 2016 atoms. We thus expect the method to be useful in a variety of situations.

This research was supported by the US National Science Foundation through TeraGrid resources provided by NCSA under grant DMR050013N, through the U.S. Department of Energy, National Energy Research Initiative for Consortia (NERI-C) grant DE-FG07-07ID14893.

\end{document}